**Magnetic-field-induced reduction of the exciton polarization splitting in InAs quantum dots**

R. M. Stevenson[1], R. J. Young[1,2], P. See[1], D. G. Gevaux[1],
K. Cooper[2], P. Atkinson[2], I. Farrer[2], D. A. Ritchie[2] and A. J. Shields[1].

[1]Toshiba Research Europe Limited, 260 Cambridge Science Park, Cambridge CB4 0WE, UK
[2]Cavendish Laboratory, University of Cambridge, Madingley Road, Cambridge CB3 0HE, UK

By the application of an in-plane magnetic field, we demonstrate control of the fine structure polarisation splitting of the exciton emission lines in individual InAs quantum dots. The selection of quantum dots with certain barrier composition and confinement energies is found to determine the magnetic field dependent increase or decrease of the separation of the bright exciton emission lines, and has enabled the splitting to be tuned to zero within the resolution of our experiments. Observed behaviour allows us to determine g-factors and exchange splittings for different types of dots.

The atom-like properties of single semiconductor quantum dots, together with their ease of integration into more complicated device structures, has made them an attractive, and widely studied system for applications in quantum information [1]. One of their potentially useful properties is the emission of pairs of polarisation entangled photons by the radiative decay of the biexciton state [2]. However, experiments have shown that the realisation of such a device is prevented by the lifting of the degeneracy of the intermediate exciton level, resulting in only classically correlated photon pair emission [3,4]. The physical origin of the lifting of the degeneracy is attributed to the exchange interaction [5], the in-plane symmetry of which is broken by the structure of the quantum dot. This results in linear polarisation splitting in the exciton and biexciton emission. The ability to control this splitting is essential in order to realise an on-demand entangled photon pair source.

Recently, time domain measurements of annealed quantum dot ensembles have demonstrated a clear reduction in the splitting [6,7], and photoluminescence of individual dots has shown that control of the size of quantum dots by growth alone is enough to nullify and even invert the splitting [8]. However, the irregular nature of quantum dot sizes and shapes results in an ensemble with a distribution of splittings, so both growth control and annealing do not provide a convenient method to tune the splitting of a single quantum dot. In this Letter, we demonstrate the effect of magnetic fields on the splitting, and describe the circumstances for which we observe a reduction in the splitting to zero.

The radiative decay of the neutral biexciton (XX) state is found experimentally to dominate quantum dot emission spectra. The biexciton state consists of a pair of electrons with spins +1/2 and -1/2, and a pair of holes with spins +3/2 and -3/2. The biexciton state (and ground state) is therefore spin neutral, and has zero net Zeeman interaction with the magnetic field, in contrast to the intermediate exciton (X) states, which also govern the polarisation of emitted photons [9]. There are four X states, each with an electron of spin ±1/2 and a hole of spin ±3/2, and characterised by their total angular momentum $m$ of +1,-1,+2 and -2. The X states $|m\rangle=|\pm1\rangle$ are optically active or 'bright', as they can radiatively recombine to emit a photon. The X states $|m\rangle=|\pm2\rangle$ are optically inactive, or 'dark'.

The X states are also modified by electron hole exchange interactions [5,9,10,14,16,17,18]. The resulting energy levels are shown schematically in Fig. 1a for a dot similar to the InAs dots studied here. The long range component of the exchange interaction is dominant in quantum dots [10], and pushes the dark states to lower energy than the bright states by $D_0$, typically a few 100µeV and independent of in-plane anisotropy. Long range exchange also splits the two bright X states by $S_0$ for dots with anisotropic electron hole overlap in the plane, which exists for all dots previously studied due to preferential elongation [11], strain [12,13], and diffusion along the [1-10] direction of the crystal. $S_0$ is typically at least several 10µeV. In-plane anisotropy also splits the dark states by $\sigma_0$, but this is sensitive to the short range exchange interaction only, which acts only within the unit cell, and $\sigma_0$ therefore is much smaller than $S_0$ or $D_0$, and we approximate it to zero [14,17,5]. The eigenstates of the dot are the symmetric and anti-symmetric bright states (|+1>+|-1>) and (|+1>-|-1>), and the symmetric and anti-symmetric dark states (|+2>+|-2>) and (|+2>-|-2>).

The radiative decay of XX can occur via either the symmetric or anti-symmetric bright X state to the ground state, emitting of a pair of horizontally (H) or vertically (V) linearly polarised photons, oriented along the [110] and [1-10] directions respectively. If an in-plane magnetic field is applied to the quantum dot, then the dark and bright states become coupled via the Zeeman interaction [5,9,10,14,16,17,18]. This partially allows optical transitions to and from the predominately dark, or 'darker' states, as indicated by dashed arrows in Fig. 1a. Each dark state mixes with just one bright state and shares a linear polarisation with this state, indicated by the size and colour of the points in Fig. 1b. The separation $D_H$ ($D_V$) between the H (V) polarised brighter and darker states, increases by the Zeeman interaction energy $g_H\mu_B B_x$ ($g_V\mu_B B_x$) in the limit of high magnetic fields, demonstrated by dashed black (grey) lines. Here $g_H$ and $g_V$ are the effective g-factors governing the interactions between the bright and dark exciton states, and are equal to ($g_{e,x}\pm g_{h,x}$), where $g_{e,x}$ and $g_{h,x}$ are the in-plane electron and hole g-factors respectively. For smaller fields, $D_H$ ($D_V$) is the resultant between the Zeeman energy and the splitting at zero magnetic field $D_{H0}$ ($D_{V0}$), and increases approximately quadratically with magnetic field. The corresponding change in the bright exciton splitting $S$ is also approximately quadratic for smaller fields, and linear for larger fields. Crucially the direction, or sign of the change in $S$ is dependent on properties of the quantum dot, in contrast to the case of magnetic fields applied normal to the sample, for which $S$ always increases due to the hybridisation of the bright states.

The bright exciton splitting $S$, approximated for small in-plane magnetic fields $B_x$, is determined by the solution of the Hamiltonian for the in-plane Zeeman interaction [5,16], and has the form of equation 1.

$$S(B_x) = S_0 + K B_x^2 + K' B_x^4 \qquad (1)$$

For our experiments, we find it sufficient to parameterise the response of a given quantum dot by the coefficient $K$ of $B_x^2$ only, since $K'$ is relatively small. The change in the optically dominant exciton level splitting with field $K$ is determined experimentally in units of $\mu eVT^{-2}$, and is related to properties of the quantum dot as shown in equation 2.

$$K = \frac{\mu^2}{D_0\left(1 - S_0^2/4D_0^2\right)} \left[ g_{e,x}g_{h,x} - \frac{S_0}{4D_0}\left(g_{e,x}^2 + g_{h,x}^2\right) \right] \qquad (2)$$

In contrast to the case for normally applied fields, the change in $S$ with magnetic field is therefore dependent on both the initial symmetry and size of the quantum dot, (represented by $S_0$ and $D_0$), and the sign of the g-factors. We describe below how this allows $S$ to be reduced to zero for certain quantum dots.

The quantum dot samples used for all measurements presented here were grown by MBE. Three differing barrier compositions were used; GaAs, $Al_{0.1}Ga_{0.9}As$, and $Al_{0.33}Ga_{0.67}As$, all of which were grown to a thickness of at least 250nm. The InAs quantum dot layer was then deposited directly on the barrier material, to a thickness corresponding to the threshold for island formation, around 1.6 monolayers. The dots were then capped by at least 250nm more barrier material. The areal quantum dot density for all samples was <1μm$^{-2}$, which together with a metal mask containing apertures of 2μm diameter fabricated on the surface allowed the isolation of individual quantum dots.

The samples were measured in a continuous flow helium cryostat operating at ~5K. CW laser excitation was provided with energy above the band gap of the barriers, and focussed onto the sample using a microscope objective lens. The same lens collimated the emission, which then passed through linear polarisation selection optics before being dispersed by a grating spectrometer, and measured with a liquid nitrogen cooled CCD camera. The cryostat was placed within the bore of a superconducting magnet with the field parallel to the plane of the sample.

In photoluminescence (PL), X emission is typically seen as a linearly polarised doublet [15,3] with a zero field splitting $S_0$ and a linear intensity dependence on laser power. XX emission has quadratic power dependence and has reversed polarisation splitting -$S_0$. Determining the average of $S$ measured from X and XX emission removed systematic errors introduced by the polarisation optics, and we estimate that $S$ is determined to a precision of ~0.5μeV. We present here only spectra from X recombination for clarity.

Due to variations in size and shape of self-assembled quantum dots, the confinement energy varies from dot to dot. The exciton extends further in the plane for dots with weaker confinement, which reduces the strength of the exchange interaction and consequently $S_0$. For the weakest confined dots, the splitting inverts, and the horizontally polarised exciton emission is lower in energy than the vertically polarised, attributed to competing directions of expansion between the electron and hole [8,12]. Thus the selection of the emission energy of a dot also determines its fine structure to a large degree. Furthermore, significant control of the emission energy is provided by varying the thickness of InAs deposited. Fig. 2 shows PL spectra from single dots *A*, *B* and *C*, which are each representative of three different sets of dots, corresponding to dots with GaAs barriers and typical $S_0$, $Al_{0.33}Ga_{0.67}As$ barriers with large $S_0$, and GaAs barriers with inverted $S_0$ respectively.

Dot *A* has an emission energy of ~1.382eV, in the absence of any applied magnetic field, and splitting $S_0$ of +22±1μeV. With the application of an in-plane magnetic field of 5T, the linear polarisation of the dominant lines remains, $S$ increases to +77±1μeV, and in addition a new line is seen to lower energy in the horizontally polarised spectrum [16], which we attribute to a darker X state described above, partially mixed with the higher energy, H polarised X state, as indicated by its polarisation character.

The lower panel of Fig. 2 shows $S$ increases non-linearly with magnetic field. This is a consistent with the magnetic field induced mixing of the H polarised bright X state with a dark X state as described above. This in turn increases the splitting $S$ between the V and H polarised brighter states.

At zero field, emission of dot $B$ is ~157meV higher in energy than dot $A$, which can be explained by a wider band gap energy due to intermixing of aluminium within the quantum dot region, and increased quantisation energy from the stronger confinement provided by the AlGaAs barriers. The polarisation splitting $S_0$ is much larger than for dot $A$, at +284±1μeV, due to stronger exchange interaction caused by better exciton confinement. At 5T the familiar darker state becomes visible to lower energy, but remarkably has the opposite polarisation to that observed for dot $A$, and is vertically polarised, the same as the lower energy brighter exciton state. Analogous to the case above, the two vertically polarised states repel each other under magnetic field, and this time the lower energy exciton state is pushed closer to the higher energy exciton state, reducing the splitting from +284±1μeV at 0T to +235±1μeV at 5T. The change in $S$ is observed more clearly when plotted as function of the magnetic field in the bottom panel, which shows an approximately quadratic reduction as a function of field. We reiterate that reduction of the polarisation splitting S in not possible for magnetic fields normal to the sample.

Dot $C$ emits at a slightly higher energy than dot A by ~21meV, and shows an inverted fine structure splitting of -16±1μeV. At high field, a new feature corresponding to the dark state appears, horizontally polarised, as for dot $A$. Crucially however, by 5T the splitting has changed sign to +31±1μeV, and the order of the polarised lines is reversed to those at zero field. This indicates that the coupling of the dark states to the lower energy bright state was sufficient to energetically tune the H polarised line through the V polarised line. A weak V polarised peak is also seen, with very similar energy to the darker H polarised state. The observation of both dark states is not common in these dots, unlike in other work [16,17], but the similar energy of the two dark states observed does not contradict the approximation of small $\sigma_0$. The lower panel shows $S$ measured as a function of applied field, revealing $S$ to reduce through 0±0.5μeV. From these measurements, it appears the linear polarisation splitting of the exciton emission is reduced below the homogeneous linewidth of the emission lines of ~1.5μeV (determined from the radiative lifetime) for magnetic fields within the range 2.7±0.1T.

For dots $A$-$C$, fits in the form of equation 1 are found to be in excellent agreement with $S$ as a function of $B_x$, as shown in fig. 2. Neglecting the term in $B_x^4$ is found to have little effect, decreasing the average correlation ratio of the fit $r$ only slightly, from 99.9% to 99.3%. This confirms the change in $S$ per square Tesla $K$ to be an appropriate measure of the field dependent response of $S$.

We have measured $K$ and the zero-field exciton polarisation splitting $S_0$ for a selection of different dots. The results are plotted in Fig. 3a. Dots with AlGaAs barriers have large positive $S_0$, and $K$ is large and negative. Dots with $Al_{0.1}Ga_{0.9}As$ barriers have a positive $S_0$, and $K\sim0$, i.e. $S$ is approximately constant as a function of magnetic field. Dots with GaAs barriers have small $S_0$, both positive and negative, and positive $K$. $|S|$ is reduced by the magnetic field for dots with $K$ and $S_0$ of opposite sign, typical of dots with $Al_{0.33}Ga_{0.66}As$ barriers, and dots with GaAs barriers and inverted $S_0$. For a large number of inverted split quantum dots with GaAs barriers, and one with $Al_{0.33}Ga_{0.67}As$ barriers, we observe the linearly polarised lines cross within the 5T maximum field of our experiments. These dots are represented within the dark shaded region of Fig. 3a. If higher magnetic fields of up to 10T were accessible, it should be possible to reduce $|S|$ to zero for a significant proportion of the dots with $Al_{0.33}Ga_{0.67}As$ barriers, found within the lightly shaded region of Fig. 3a.

In order to understand why it is possible to tune the exciton polarisation splitting $S$ to zero for specific groups of dots, we consider the sign and magnitude of the change in $S$ with field ($K$) in relation to the value of $S$ at 0T ($S_0$). From inspection of equation 2, $K$ is determined by $S_0$, $g_{e,x}$, $g_{h,x}$, and $D_0$, the average splitting between the bright and dark states at 0T. In general $D_0$, $g_{e,x}$ and $g_{h,x}$ can only be uniquely determined if all four X states can be observed, which is not possible for the dots studied here. However estimations can be made as follows for those dots where $K$ is large enough to allow the observation of at least one predominantly dark state.

We estimate $D_0$ for each quantum dot by measuring the darker-brighter splitting in each polarisation, $D_V$ and $D_H$, as a function of $B_x$, and extrapolating the average separation at 0T. The confinement energy $E_c$ is then measured by determining the energy of the exciton emission relative to energy of the wetting layer peak. $D_0$ is plotted as a function of the confinement energy $E_c$ for different quantum dots in Fig. 3b, which shows a clear increase by a factor of ~5 of $D_0$ with $E_c$. This is attributed to the strengthening of exchange interaction as the confinement of the exciton increases. For dots with $Al_{0.33}Ga_{0.67}As$ barriers, a similar correlation to Fig. 3b is observed when $D_0$ is plotted against $S_0$ as shown in Fig. 3c. This is expected as $S_0$ is known to increase with stronger confinement $E_c$ [8]. The ratio $S_0/D_0$ is on average 0.6±0.14. For dots with GaAs barriers, and especially for those which can be tuned to S=0, $S_0/D_0$ is small.

As a result, for dots with GaAs barriers, the sign of $K$ is determined by the product $g_{e,x} g_{h,x}$. As we measure $K$ to be positive we deduce $g_{e,x}$ and $g_{h,x}$ must be of the same sign. Under the current approximation of small $\sigma_0$, the weak mixing of the darker V polarised state suggests that $g_V = g_{e,x} - g_{h,x} \sim 0$, and therefore $|g_{e,x}| \sim |g_{h,x}|$. For these dots $g_H$ ($=g_{e,x}+g_{h,x}$) was determined by fitting $D_V$ as function of $B_x$, and the average was found to be 0.79±0.19. We therefore estimate $|g_{e,x}| \sim |g_{h,x}| \sim \pm 0.4 \pm 0.1$. With average $D_0$ of 215±45µeV, this corresponds to $K$=2.5±1.8, which agrees with the range of $K$ values measured. This indicates that the crossing of the brighter exciton states is due to weak coupling of the V polarised state due to similar electron and hole g-factors.

For dots with $Al_{0.33}Ga_{0.67}As$ barriers, the average $g_V$ was 1.08±0.19. To obtain the observed average $K$ of -1.67±0.94 requires $g_{e,x}=\pm 1.21$ and $g_{h,x}=\mp 0.13$, assuming that as for quantum wells, $g_{h,x}$ is smaller than $g_{e,x}$. The order of magnitude difference in the electron and hole g-factors is sufficient to allow the term in $-g_e^2$ to dominate $K$, originating from stronger mixing of the energetically closer, lower energy bright state with the dark states, and resulting in the reduction of S with increasing magnetic field.

In summary, we have demonstrated that for many dots it is possible to engineer a crossing of the typically non-degenerate exciton levels by the application of modest magnetic fields in the plane of the sample. We conclude that the magnetic field response of $S$ is strongly dependent on the 0T fine structure, and $g$ factors of each dot. Two types of quantum dot are identified for which linear polarisation splitting can be tuned to zero by the application of an in-plane magnetic field. Suitable dots can be selected by the choice of the barrier material, and the the emission energy of the dot, which can is related to the InAs deposition thickness. The first type, dots with GaAs barriers and inverted initial polarisation splitting, are easily tuned to $S$=0 by modest fields, a consequence of small negative $S_0$ and similar electron and hole g-factors. The second type, dots with AlGaAs barriers and large $S_0$, require stronger fields due to the larger $D_0$, and $S_0$, and are dominated by the effects of the larger, electron g-factor.

It is important to consider the effect of the hybridisation of the dark and bright states in terms of entangled photon pair emission from exciton states magnetically tuned onto resonance as described here. At least for the dots presented here, one of the bright states tends to couple more strongly to the dark states, which thus preferentially increases its radiative lifetime due to the inhibited recombination of the dark state component [18]. In principle it is therefore possible to distinguish the predominantly bright mixed states in the time domain. In practice however, the change to the radiative lifetime, and consequently the homogeneous linewidth, is small for the weakly coupled states presented here, and consequently one would expect that entangled photon emission may well be possible, in the absence of any interactions hidden within the resolution limit of these experiments.

The authors would like to acknowledge financial support from the EU projects RAMBOQ, QAP and SANDiE, and the EPSRC through the IRC for Quantum Information Processing.

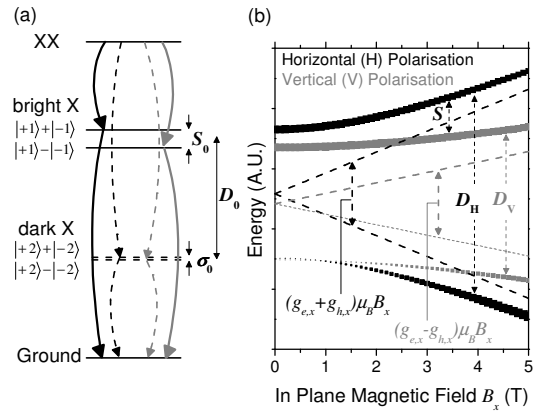

Fig. 1 (a) Energy level schematic of a single quantum dot. (b) Fine structure of the exciton state (X) in a single quantum dot as a function of in-plane magnetic field $B_x$. The size of the black (grey) points represents the fraction of horizontal (vertical) polarisation. The areas of the points corresponding to darker states has been multiplied by 5 for clarity.

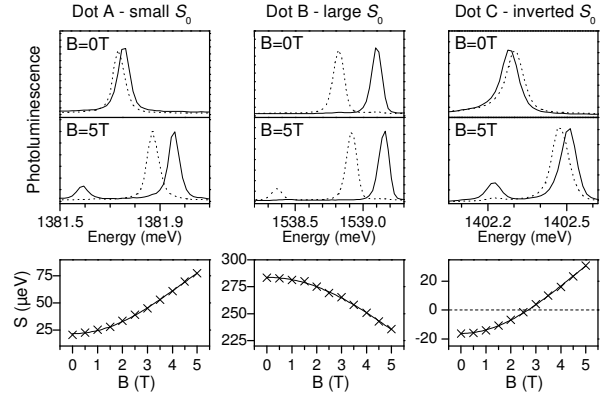

Fig. 2. Top panels show vertically and horizontally polarised photoluminescence spectra as dashed and solid lines respectively for the neutral exciton states of dots *A*, *B*, and *C* defined in the text. Middle panels show the same with an in-plane magnetic field of 5T. Bottom panels show separation of dominant horizontally polarised emission line relative to dominant vertically polarised emission line as a function of magnetic field. Lines show fit to observed behaviour.

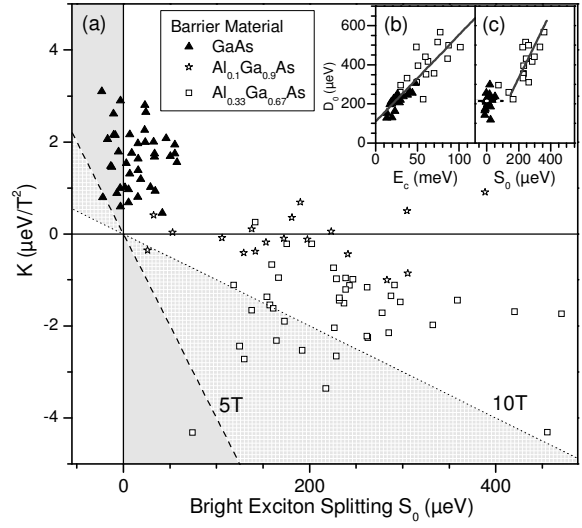

Fig. 3. (a) Coefficient $K$, characterising the change in optically dominant exciton splitting $S$ as a function of magnetic field, as a function of the splitting at zero field $S_0$, for different dots. Shaded regions show dots for which brighter exciton states cross for fields below 5 and 10T. (b) Extrapolated Bright-dark exchange splitting $D_0$ as a function of the confinement energy $E_c$ for different quantum dots. Line shows linear best fit to data. (c) $D_0$ as a function of $S_0$. Solid line show best fit linear dependence for dots with $Al_{0.33}Ga_{0.67}As$ barriers. Dashed line indicates average $D_0$ for dots with GaAs barriers.